\documentclass[aps,twocolumn,amsmath,amssymb,preprintnumbers]{revtex4}
\usepackage{amsmath} \usepackage{amsfonts} \usepackage{amssymb}
\usepackage{bbm}
\usepackage{epsfig}
\usepackage{graphics}
\usepackage{graphicx}
\usepackage{hhline}
\usepackage{booktabs}
\usepackage{array}
\usepackage{caption}
\newcommand{\be}{\begin{equation}}
\newcommand{\ee}{\end{equation}}
\newcommand{\ba}{\begin{eqnarray}}
\newcommand{\ea}{\end{eqnarray}}
\newcommand{\nn}{\nonumber}

\newcommand{\mn}{_{\mu\nu}}

\begin{document}

\title[ ]{Cosmon inflation}

\author{C. Wetterich}
\affiliation{Institut  f\"ur Theoretische Physik\\
Universit\"at Heidelberg\\
Philosophenweg 16, D-69120 Heidelberg}

\begin{abstract}
The cosmon field responsible for dynamical dark energy can induce an inflationary period in early cosmology. Details depend on dilatation symmetry breaking terms in its potential and kinetic term that are relevant only during early stages of the evolution of the universe.  
For a large range of parameters we find realistic primordial density fluctuations. As a particular example we
discuss a model with almost static geometry in the Jordan frame for all epochs of cosmology. Here the cosmic evolution of the Hubble parameter is replaced by a variation of the Planck mass and particle masses. This model predicts a spectral index $n=0.97$ and a ratio for tensor to scalar fluctuations $r=0.13$.
\end{abstract}

\maketitle
An inflationary period \cite{Inf1,Inf1A,Inf2,Inf3,Inf4,Inf5} in very early cosmology and the dominance of dark energy in very late cosmology are central elements of our present understanding of the universe. They are tested jointly by anisotropies in the cosmic microwave background (CMB) \cite{WMAP7}, \cite{PL}. Both the very early and the very late stage of the cosmological evolution look actually very similar: the energy density of the universe is dominated by an almost static component with equation of state $p\approx -\rho$. Inflation is usually described by a scalar field, the inflaton. Here the necessity of a dynamical field arises since the inflationary period has to end. The proposal of dynamical dark energy \cite{CW3} is also based on the evolution of a scalar field, the cosmon. An obvious question asks if the inflaton and the cosmon could be the same field $\chi$. In principle, this is well possible, as argued in refs. \cite{PV}, \cite{PX}.

 Inflation and dark energy domination are related to very different epochs in cosmology and correspond to different values of the scalar field $\chi$. The properties of the effective action for $\chi$, i.e. potential and kinetic term, typically depend on the value of $\chi$ and can therefore be quite different for the two epochs. Based on this, one may taylor an action with suitable properties. It should account for (i) inflation, (ii) entropy production after inflation, (iii) a subdominant scalar energy density during the radiation and matter dominated epochs, (iv) domination of the energy density by the cosmon in the recent cosmological epoch. A good candidate for this sequence of epochs is a scaling solution \cite{CW3}, \cite{RP}, \cite{SL1}, \cite{SL2}  for the radiation and matter dominated period. For such a scaling solution the ratio of scalar energy density $\rho_h$ to the dominant radiation or matter component $\rho_r$ or $\rho_m$ remains constant. Thus $\rho_h$ can be associated with ``early dark energy'' \cite{EDE}, with a fixed fraction $\Omega_h=\rho_h/\rho_c$ of perhaps a few percent. The long duration of the scaling, where $\rho_h$ decreases with the inverse second power of time, can explain the small amount of the present value of $\rho_h$ in units of the Planck mass, $\rho_h/M^4\approx 10^{-120}$. 

In this scenario we can divide the history of the universe in four periods. A first scalar field dominated period (inflation) is followed by the radiation and matter dominated periods for which the scalar field is subdominant and follows a scaling solution. Finally, a new scalar field dominated period has begun rather recently. All three transitions in this sequence need some reason for the end of the preceding epoch. For inflation it is typically the end of the ``slow roll'' property. The end of radiation domination is triggered by matter becoming more important since it is diluted more slowly by the cosmic expansion. Finally, the end of the matter dominated period can be associated to an effective stop of the evolution of the scalar field. This results in constant $\rho_h$ and ends the scaling solution, such that $\rho_h/\rho_m$ increases. The stop of the scalar field may result from a qualitative change of the potential or kinetic term for a characteristic value of $\chi$. Alternatively, it could be induced by a cosmological trigger event, as neutrinos becoming non-relativistic in growing neutrino quintessence \cite{ABW,CWNEU} or by the formation of non-linear structures. 

For the description of dynamical dark energy by the cosmon the symmetry of dilatations or scale transformations plays an important role. The cosmon is the ``pseudo-Goldstone boson'' of spontaneously broken dilatation symmetry. Its potential arises from effects breaking the dilatation symmetry explicitly (``dilatation anomaly''). During the scaling solution these effects become less and less important. For infinite time a fixed point with exact dilatation symmetry is approached. At the fixed point, the cosmon would be an exactly massless Goldstone boson. We can now move backwards in cosmology. Then the scale symmetry breaking terms in the action become more and more important. They may therefore play a crucial role in very early cosmological epochs as inflation. 

In this note we demonstrate that scale symmetry breaking terms in the scalar potential can indeed be responsible for an inflationary period. They are associated with a characteristic intrinsic mass scale $\mu$. (Parameters in the action with dimension of mass or length break the scale symmetry explicitly.) A second important effect is a shift of the effective kinetic term for the cosmon at some characteristic scale $\bar m$. We investigate models where the coefficient of the kinetic term $\partial^\mu\chi\partial_\mu\chi$ is large for $\chi\ll \bar m$, and close to the conformal value for $\chi\gg \bar m$. The end of inflation will typically occur near the transition between the two regimes at $\chi\approx \bar m$. Details of the inflationary period depend, in general, on the detailed structure of the scale symmetry breaking terms. 

As an example we discuss a class of very simple models with a scalar potential $\tilde V=\mu^2\chi^2$, formulated in the Jordan frame with effective Planck mass depending on $\chi$. For this model the curvature scalar $\tilde R$ stays almost constant for the whole history of the universe, $\tilde R\approx \mu^2$, with $\mu=2\cdot 10^{-33}$eV. Only the effective Planck mass increases, being proportional to $\chi$ for
large values of $\chi$ which characterize
 the ``late'' cosmological epochs after inflation. Also particle masses scale proportional to $\chi$. 
For constant
 dimensionless ratios of masses and dimensionless couplings all bounds from time variation of fundamental constants or apparent violations of the equivalence principle are obeyed. We find that this simple class of models is quite predictive for the slow roll parameters of inflation. It predicts for the spectral index $n=0.97$ and for the ratio of tensor to scalar fluctuations $r=0.13$. 

We also discuss more general models, for example by adding in the Jordan frame a cosmological constant. We find  that the details of the primordial density fluctuations depend on the parameters of the model, often with ranges $0.02<r<0.08,\ 0.93<n<0.96$. Such models seem to be compatible with the present observational knowledge.

\medskip\noindent
 {\em Effective action.}
\smallskip

 The quantum effective action for the cosmon-graviton system is assumed to take the form 
\be\label{1}
S_{cg}=\int_x\sqrt{\tilde g}\left\{\frac12 K(\chi)\partial^\mu\chi\partial_\mu\chi+\tilde V(\chi)-\frac12 F(\chi)\tilde R\right\},
\ee
with $\tilde R$ the curvature scalar of the metric $\tilde g\mn$ in the Jordan frame. We investigate simple polynomial forms of 
$\tilde V$ and $F$,
\ba\label{2}
\tilde V(\chi)&=&\mu^2\chi^2+\bar\lambda_c,\\
F(\chi)&=&\chi^2+m^2.\label{3}
\ea
For large $\chi^2$ we can neglect $m^2$ in $F$ such that $\chi$ plays the role of a variable (reduced) Planck mass \cite{CW1}. This region will be reached for late cosmology. With the potential \eqref{2} and for suitable $K$ this entails a quintessence cosmology \cite{CW3}, see below. In contrast, for small $\chi^2$ the effective Planck mass and the potential become independent of $\chi$. This region is assumed to dominate the very early stage of cosmology. For all $\chi$ we may associate $F^{-1}(\chi)$ with an effective gravitational constant. Higher order terms in an expansion of $m^2/F(\chi)$ or $\mu^2/F(\chi)$ could be added to $\tilde V(\chi)$. We omit them because they do not change the qualitative picture. The asymptotic form $F\rightarrow\chi^2$ is required for asymptotic dilatation symmetry. (A multiplicative constant can be absorbed in the definition of $\chi$.) Thus eq. \eqref{3} is a simple form connecting asymptotic scale symmetry with a constant for small $\chi$. The dimensionful parameters $\mu^2$, $\lambda_c$ and $m^2$ violate scale symmetry.

 For the cosmon kinetic term we choose 
\be\label{3A}
K(\chi)=4\omega+6\tau\frac{m^2}{F(\chi)},
\ee
omitting again higher orders in $m^2/F$. For $\chi\to \infty$ one has $K(\chi)\to 4\omega$, with $\omega$ corresponding to a similar parameter in a Brans-Dicke theory \cite{BD}. Stability for large $\chi$
requires $\omega>-3/2$, with $\omega=-3/2$ singled out as the ``conformal point''. We choose $\tau>-2\omega/3$ in order to guarantee stability also for finite $\chi$ and negative $\omega>-3/2$. The ansatz \eqref{3A} is a simple form interpolating between two different constants for large and small $\chi$, namely $4\omega$ and   $4\omega + 6\tau$. The precise form of this interpolation does actually not matter for the qualitative results of this note. 

The field equations derived by variation of the effective action \eqref{1} show some unusual features due to the $\chi$-dependence of the coefficient of the curvature scalar $\tilde R$. In particular, the cosmological value $\chi(t)$ can {\em increase} with time despite the fact that the minimum of the effective potential $\tilde V(\chi)$ occurs at $\chi=0$ \cite{CW1,CW3}. 

\medskip\noindent
 {\em Einstein frame.}
\smallskip

The quantitative evolution is followed most easily if we perform a field transformation from the ``Jordan frame'' to the ``Einstein frame''. For this purpose we employ a rescaling of the metric (for details see ref. \cite{CW1})
\be\label{4}
\tilde g\mn= w^2 g\mn~,~w^2=M^2/F(\chi).
\ee
Expressed in the new ``field coordinates'' $g_{\mu\nu}$ the action reads
\be\label{5}
S_{cg}=\int \sqrt{g}\left\{\frac{2M^2k^2(\chi)}{\chi^2}\partial^\mu\chi\partial_\mu\chi+V(\chi)-\frac{M^2}{2}R\right\},
\ee
with
\be\label{6}
V(\chi)=\frac{M^4\tilde V(\chi)}{F^2(\chi)}=
\frac{M^4(\mu^2\chi^2+\bar\lambda_c)}{(\chi^2+m^2)^2}
\ee
and 
\ba\label{7}
k^2(\chi)&=&\frac{\chi^2}{4}\Big\{\frac KF+\frac32\left(\frac{\partial \ln F}{\partial \chi}\right)^2\Big\}\nn\\
&=&\frac{\chi^2}{\chi^2+m^2}\left\{\omega+\frac32+\frac32(\tau-1)\frac{m^2}{\chi^2+m^2}\right\}.
\ea

For large $\chi$ one has
\ba\label{8}
\lim_{\chi\to\infty}V(\chi)&=&\frac{M^4\mu^2}{\chi^2},\nn\\
\lim_{\chi\to\infty}k^2(\chi)=k^2_\infty&=&\omega+\frac32=\frac{1}{\alpha^2}.
\ea

In the Einstein frame the potential vanishes for $\chi\to\infty$, corresponding to an ``asymptotically vanishing cosmological constant'' \cite{CW2}. We choose $M$ to be equal to the present value of the scalar field $\chi_0=\chi(t_0)=\chi(a=1)$ and associate it with the present reduced Planck mass, $M^2=8\pi G_N,M=2.44\cdot 10^{27}$eV. If we identify $V(\chi_0)$ with the present dark energy density we obtain
\be\label{9}
\frac{V(\chi_0)}{M^4}=\frac{\mu^2}{M^2}\approx 7\cdot 10^{-121}~,~\mu=2\cdot 10^{-33}{\rm eV}.
\ee
This implies that $\mu$ is of the order of the present Hubble parameter $H_0$. Setting the scale $\mu$ is actually arbitrary - we could set $\mu=1$ by an appropriate rescaling of fields. The only
dimensionless parameters appearing in the action  \eqref{1} are $\omega,\ \tau $ and the ratios $m/\mu,\ \bar\lambda_c/\mu^4$. The tiny 
  ratio $\mu^2/\chi^2(t_0)$ is a consequence of the increase of $\chi$ for a long cosmological period. Indeed, $\mu^2/\chi^2(t)$
can become very small if $\chi(t)$ increases sufficiently fast and for a sufficient time. This 
scenario is realized in the present
 type of models.

Defining $(\mu>0)$ 
\be\label{10}
\varphi=2M\ln(\chi/\mu)
\ee
the scalar part of the action becomes for large $\chi$
\be\label{11}
S_c=\int_x\sqrt{g}\left\{\frac12 k^2(\varphi)\partial^\mu\varphi\partial_\mu\varphi+M^4\exp \left\{-\frac{\varphi}{M}\right\}\right.,
\ee
with ``kinetial'' $k^2(\varphi)$ approaching the constant \eqref{8} for $\varphi\to\infty$. One recovers quintessence with an exponential potential, which shows for the dominance of matter $(n=3)$ or radiation $(n=4)$ the well known scaling solution \cite{CW2} with a constant early dark energy fraction 
\be\label{11A}
\Omega_h=nk^2_\infty=\frac{n}{\alpha^2}.
\ee
Observational constraints on early dark energy require $k_\infty\lesssim 0.1$ \cite{A2a,A2b,Re,A2c,A2d}. Realistic cosmology needs an end of this scaling solution around redshift $z\approx 5$, for example as a consequence of a cosmological trigger event as for growing neutrino quintessence \cite{ABW,CWNEU}. Such a trigger event  stops the further time evolution of the cosmon $\chi$. It typically involves additional ``late time scaling violation'' besides the parameter $\mu$, as for the case of neutrino masses growing faster than $\chi$.

\medskip\noindent
 {\em Slow roll approximation for inflation.}
\smallskip

We are mainly interested here in the early inflationary period of cosmology for not too large $\chi$. In this region of the field space the cosmon is associated with the inflaton. In order to be close to the usual normalization we employ
\be\label{11B} 
\tilde\sigma=\frac{2M}{m}\chi,
\ee
such that for $\chi\approx m$ one has $\tilde \sigma\approx M$. With this normalization the scalar part of the action \eqref{5} reads
\be\label{12}
S_c=\int_x\sqrt{g}\left\{\frac12 Z(\tilde\sigma)\partial^\mu\tilde \sigma\partial_\mu\tilde \sigma+V(\tilde\sigma)\right\},
\ee
with 
\ba\label{13}
V&=&\gamma M^4\left(1+\frac{x}{\beta}\right)(1+x)^{-2},\nn\\
Z&=&\frac {k^2}{x}=\frac{\omega}{1+x}+\frac{3(\tau +x)}{2(1+x)^2}.
\ea
Here we have introduced the dimensionless variables and parameters
\ba\label{14}
x&=&\frac{\tilde \sigma^2}{4M^2}=\frac{\chi^2}{m^2}~,\\
\gamma&=&\frac{\bar\lambda_c}{m^4}~,~\beta=\frac{\gamma m^2}{\mu^2}=\frac{\bar\lambda_c}{m^2\mu^2}.\nn
\ea
For small values of $\gamma$ the potential can be substantially smaller than $M^4$.

The canonically normalized scalar field $\sigma$ is related to $\tilde \sigma$ by
\be\label{15}
\frac{\partial\sigma}{\partial\tilde \sigma}=Z^{1/2}(\tilde\sigma).
\ee

With
\be\label{15A}
\frac{\sigma}{M}=\int_{0}^{x}dx'\sqrt{\frac{Z(x')}{x'}}         
\ee
this defines an implicit equation for $V(\sigma)$. In the region of interest for inflation (horizon crossing) one typically has $x\ll \tau$. With $\omega\approx -3/2$ one infers for this region
\be\label{18B}
Z=\frac{3(\tau-1)}{2(1+x)^2},
\ee
such that
\be\label{18C}
\frac{\sigma}{M}=\sqrt{\frac{3}{2}(\tau-1)}\int dx'\frac{1}{\sqrt{x'(1+x')}}.
\ee
In the region of small $x\ll1$ this yields a quadratic relation
\be\label{18D}x=\frac{1}{3(\tau-1)} \frac{\sigma^2}{M^2},
\ee
while for $x\gg 1$ one has an exponential dependence
\be\label{18E}
x\sim\exp\left(\sqrt{\frac{2}{3(\tau-1)}}\frac{\sigma}{M}\right).
\ee
Insertion into eq. \eqref{13} yields the potential for the normalized inflaton field $\sigma$.

We can determine the slow roll parameters of inflation in the usual way. One finds
\ba\label{16}
\epsilon&=&\frac{M^2}{2}\left(\frac{\partial\ln V}{\partial\sigma}\right)^2=\frac{x}{2Z}
\left(\frac{\partial\ln V}{\partial x}\right)^2\nn\\
&=&\frac{x}{2\omega(1+x)+3(\tau+x)}\left(2-\frac{1+x}{\beta+x}\right)^2.
\ea
For $\beta<1/2$ one has a maximum of $V$ at
\be\label{16A}
x_{\rm max}=1-2\beta,
\ee
which corresponds to $\epsilon(x_{\rm max})=0$. In this case the initial value of $x$ should be larger then $x_{\rm max}$. For the sake of simplicity we mainly concentrate on $\beta \geq 1/2$, keeping in mind that $\beta=1/2$ is an interesting point in parameter space where the potential around $x=0$ is particularly flat. The second slow roll parameter obtains as
\ba\label{17}
\eta&=&\frac{M^2}{V}\frac{\partial^2 V}{\partial\sigma^2}=2\epsilon+M^2\frac{\partial^2\ln V}{\partial\sigma^2}\nn\\
&=&2\epsilon+\left(\frac{\partial\ln V}{\partial x}\right)^{-1}\frac{\partial\epsilon}{\partial x}\nn\\
&=&2\epsilon-\left(\frac{2}{1+x}-\frac{1}{\beta+x}\right)^{-1}\frac{\partial\epsilon}{\partial x}.
\ea
It is typically of a similar size as $\epsilon$. It's sign may, however, be negative.  

For $\beta \geq 1/2$ the maximum of $V$ is at $x=0$. In the region around this maximum $\epsilon$ is always small, 
\be\label{18}
\epsilon=\frac{4x}{2\omega+3\tau}\left(\frac{\beta-\frac12}{\beta}\right)^2,
\ee
while $\eta$ obeys
\be\label{19}
\eta=-\frac{2}{2\omega+3\tau}\frac{\beta-\frac12}{\beta}.
\ee
Thus also $\eta$ is small and negative provided $\tau$ is large enough or $\beta$ is close to $1/2$. (For $\beta< 1/2$ a similar situation arises close to the maximum of $V$ at $x_{\rm max}$.) For simplicity we may take $\tau$ large enough such that the slow roll condition is always fulfilled for $x\lesssim 1$. There is no problem to obtain a sufficient duration of inflation. On the other hand, for very large $x$ one has
\be\label{20}
\epsilon=\frac{1}{2\omega+3}=\frac{1}{2k^2_\infty}
\ee
Since $k^2_\infty$ must be small according to eq. \eqref{11A} we conclude that the slow roll region has to end before the asymptotic behavior for $x\to\infty$ is reached. Typically, the end of the inflationary period may be situated around $x_f$ with $\epsilon(x_f)\approx 1$.

\medskip\noindent
 {\em Primordial density fluctuations.}
\smallskip

In the slow roll approximation the amplitude of scalar density fluctuations generated during the inflationary phase obeys
\be\label{21}
\Delta^2=\frac{1}{24\pi^2\epsilon}\frac{V}{M^4},
\ee
where $V$ and $\epsilon$ have to be evaluated for the value of $\sigma$ at the time when the characteristic scale leaves the horizon. This happens around 50-60 $e$-foldings before the end of inflation, depending on the details of the subsequent heating of the Universe. A realistic size of the fluctuations requires $\Delta^2=2.95\cdot 10^{-9}A,A\approx 0.7$.

The ratio of tensor to scalar fluctuations is given by
\be\label{22}
r=16\epsilon.
\ee
Finally, the spectral index
\ba\label{23}
n=1-6\epsilon+2\eta
\ea
is found by observation \cite{PL} as $n\approx 0.96$. Here $\epsilon$, $\eta$, $r$ and $n$ have to be evaluated when fluctuations leave the horizon. The number of e-foldings before the end of inflation obeys

\be\label{24}
N(\bar x)=\int^{x_f}_{\bar x}\frac{Z}{d\ln V/ d\ln x}dx=\int^{x_f}_{\bar x}\sqrt{\frac{Z}{2x\epsilon}}dx.
\ee

With eqs. \eqref{16}, \eqref{17}, \eqref{24} and $\epsilon(x_f)=1$ we can compute $r$ and $n$ according to eqs. \eqref{22}, \eqref{23}. They depend on the dimensionless parameters $\omega$, $\tau$ and $\beta$. We fix $\omega+\frac32=k^2_\infty=0.01$ and show in tables I, II the values of $n$ and $r$ for several combinations of $\tau$ and $\beta$, using $N(\bar{x})=60$. The amplitude of density fluctuations further involves $\gamma$ and we display the value of $\gamma$  required for a realistic amplitude also in tables I, II. We also indicate in the tables the value of $m/\mu=\sqrt{\beta/\gamma}$ and the value $\bar{x}$ when density fluctuations leave the horizon. 
{\renewcommand{\arraystretch}{2}%
\begin{table}[h]
\captionsetup{font=scriptsize}
 \begin{tabular}[c]{|c|c|c|c|c|}
\hline $\tau$ & 15 & 20 & 500 & $10^4$ \\ \hhline{=====}
$n$ & 0.95 & 0.96 & 0.97 & 0.97 \\ \hline
$r$ & 0.02 & 0.04 & 0.13 & 0.13\\ \hline
$\gamma$ & $7.4\cdot10^{-10}$ & $1.3\cdot10^{-9}$ & $5\cdot10^{-8}$ & $10^{-6}$ \\ \hhline{=====}
$m/\mu$ & $3.7\cdot10^5$ & $27000$ & $4500$ & $1000$ \\ \hline
$\bar{x}$ & $0.06$ & $0.14$ & $12$ & $248$ \\ \hline
\end{tabular}\medskip

{\footnotesize Table I: Properties of primordial density fluctuations for $\beta=1$}
\label{Tab1}
\end{table}

\begin{table}[h]
\captionsetup{font=scriptsize}
 \begin{tabular}[c]{|c|c|c|c|c|}
\hline $\tau$ & 20 & 30 & 40 & 100 \\ \hline\hline
$n$ & 0.93 & 0.945 & 0.95 & 0.955 \\ \hline
$r$ & 0.016 & 0.05 & 0.08 & 0.17\\ \hline
$\gamma$ & $5.2\cdot10^{-10}$ & $1.7\cdot10^{-9}$ & $3.2\cdot10^{-9}$ & $1.8\cdot10^{-8}$ \\ \hhline{=====}
$m/\mu$ & $4.4\cdot10^6$ & $2.4\cdot10^6$ & $1.8\cdot10^6$ & $7.5\cdot10^5$ \\ \hline
$\bar{x}$ & $0.015$ & $0.07$ & $0.15$ & $0.83$ \\ \hline
\end{tabular}\medskip

{\footnotesize Table II: Properties of primordial density fluctuations for 
$\beta=100$}
\label{Tab2}
\end{table}
}
We observe that the small size of the amplitude of the density fluctuations requires some of the dimensionless quantities to be substantially larger or smaller than one, but no fine tuning of parameters is needed (see below). 
In the cosmon sector the parameters $\omega$, $\tau$, $\gamma$ and $\beta$ are the only parameters of the model - we recall that the present value $\mu/\chi=\mu/M$ is dynamical and tiny as a consequence of the huge age of the Universe.

\medskip\noindent
 {\em Universal large $x$-limit.}
\smallskip

We next turn to an interesting limit where the properties of primordial density fluctuations become rather universal, i.e. requiring not much information about the detailed model parameters. For
 large $\tau$ the relevant last phase of inflation occurs for $x\gg 1$.  In this limit one has 
\be\label{25}
Z=\frac{3\tau}{2x^2}+\frac{k^2_\infty}{x}
\ee
and
\be\label{26}
\epsilon=\frac{D_\beta}{2Zx}=\frac{D_\beta x}{3\tau+2k^2_\infty x}.
\ee
The factor $D_\beta$ depends on $x/\beta$, 
\be\label{27}
D_\beta=\Big(2-\frac{1+x}{\beta+x}\Big)^2,
\ee
and varies between one for large $x/\beta$ and four for small $x/\beta$.
For a simple analytic discussion one may take for $D_\beta$ some suitably averaged constant value between one and four. 

With the condition $\epsilon(x_f)=1$ inflation ends for 
\be\label{28}
x_f=\frac{3\tau}{D_\beta(x_f)-2k^2_\infty},
\ee
consistent with $x_f \gg 1$ for $\tau\gg 1$. Furthermore, with eqs. \eqref{24}, \eqref{25}, \eqref{27} the association between $x$ and the number of $e$-foldings before the end of inflation $N(x)$ becomes
\ba\label{29}
N(x)&=&\frac{1}{\sqrt{D_\beta}}\int^{x_f}_x Z(x)dx\nn\\
&=&\frac{1}{{\sqrt D_\beta}}\left\{\frac{3\tau}{2}
\left(\frac1x-\frac{1}{x_f}\right)+k^2_\infty \ln\frac{x_f}{x}\right\}.
\ea
For $N(\bar x)=60$ one finds
\be\label{30}
\bar x\approx \frac{\tau}{40\sqrt{D_\beta}}\approx \frac{\sqrt{D_\beta}}{120}x_f,
\ee
such that during this last phase of inflation $x$ increases by about two orders of magnitude from $\bar x$ to $x_f$. The validity of the approximation $\bar x\gg 1$ requires $\tau\gg 100$ and corresponds to the last two entries in  table I.

We can now extract the relevant parameters for the density fluctuations. Inserting the value \eqref{30} one finds
\ba\label{31}
\epsilon&=&\frac{\sqrt{D_\beta}}{120}~,~r=\frac{2\sqrt{D_\beta}}{15},\nn\\
\eta&=&2\epsilon-\frac{1}{120}~,~n=1-\frac{1}{60}(\sqrt{D_\beta}+1), 
\ea
while the observed amplitude obtains for 
\be\label{32}
\frac{V(\bar x)}{M^4}\approx 4\sqrt{D_\beta}10^{-9}.
\ee
We distinguish two cases: For $\bar x\gg\beta$ one has $D_\beta=1$, (case A), while for $\bar x\ll \beta$ one finds $D_\beta=4$ (case B). The observed value of $\Delta^2$ requires then 
\be\label{32A}
(A):\quad \frac\gamma\beta=\frac{\mu^2}{m^2}\approx 10^{-10}\tau,
\ee
or
\be\label{33}
(B):\quad \gamma=\frac{\bar\lambda_c}{m^4}\approx \frac18\cdot 10^{-12}\tau^2.
\ee
The transition between the two regimes is located around $\beta\approx \tau/80$, with (A) realized for $\beta\ll \tau/80$ and (B) for $\beta\gg\tau/80$. We observe that the fluctuation amplitude is determined by the relative size of two scale-violating parameters: The scale violation in the potential is determined by $\mu^2(A)$ or $\bar\lambda_c(B)$, while the scale violation in the scalar kinetic term involves the parameter $\tau m^2$. Realistic amplitudes for density fluctuations are found for a comparatively small scale violation in the potential, $\mu^2/\tau m^2\ll 1$ of $\bar\lambda_c/(\tau m^2)^2\ll 1$. This corresponds to a rather generic situation and does not need fine-tuning of parameters.

For the case (A) a good approximation takes $\bar\lambda_c=0$ and omits the term $m^2$ in $F(\chi)$. During the inflationary period the parameter $k_\infty$ or $\omega$ only mildly influences the end of the inflationary period, while it only induces tiny corrections for the properties of density fluctuations. The latter depend only on one dimensionless ratio, $\mu^2/(\tau m^2)$, which fixes the amplitude. The slow roll parameters are independent of this ratio, such that this scenario is rather predictive,
\be\label{34}
(A):\quad n-1=-0.033~,~r=4(1-n)=0.13.
\ee
The situation is analogous for the scenario (B) which is also predictive
\be\label{35}
(B):\quad n-1=-0.05~,~r=0.27.
\ee
The transition between (A) and (B) interpolates between the values \eqref{34} and \eqref{35}. We conclude that the limit $\bar{x} \gg 1$ of our model predicts rather substantial primordial gravitational waves. The bounds from Planck data \cite{PI}, which have been published after the first version of this note, disfavor the scenario (B), while scenario (A) may be considered as borderline if the nonzero neutrino masses and the Early Dark Energy present in our model are taken as additional free parameters. 

\medskip\noindent
 {\em Model with almost constant curvature.}
\smallskip

As an instructive example of our model we may consider the parameters $\tau=10^6,\beta=0.5,m=10^2\mu$ for which (A) is a very good approximation. It is instructive to discuss this model in the Jordan frame, eqs. \eqref{1}-\eqref{3A}. For both the inflationary phase and the subsequent scaling solution the potential \eqref{2} is given by a simple ``mass term'' $\tilde V=\mu^2\chi^2$. The inflationary phase and the scaling solution are only distinguished by the effective kinetic term for the cosmon. For the inflationary phase one has $K=12\tau m^2/\chi^2$, while soon after the end of inflation the kinetic term is dominated by $K=8\omega$, cf. eqs. \eqref{3A}, \eqref{25}. Most interestingly, the Hubble parameter $\tilde H$ in the Jordan frame remains essentially constant for the whole cosmological evolution. Indeed, for a slow evolution of the cosmon the curvature scalar $\tilde R=12\tilde H^2$ obeys
\be\label{36}
\tilde R=\frac{4\tilde V}{\chi^2}=4\mu^2.
\ee
This constant value is only slightly modified in the presence of radiation or matter. Indeed, for the scaling solution one has now
\be\label{37}
\tilde R=\frac{8\tilde V}{\Omega_h(1-w_h)\chi^2}=\frac{8\mu^2}{\Omega_h(1-w_h)},
\ee
with $\Omega_h=nk^2_\infty$ the fraction of (early) dark energy and $w_h$ its equation of state. Also in the present dark energy dominated epoch eq. \eqref{37} remains valid, now with $\Omega_{h,0}\approx 0.7,w_{h,0}\approx -1$. We arrive at a remarkable new picture of cosmology. It is not the geometry that changes, but only the strength of the gravitational interaction. The very early inflationary cosmology and the present universe differ only by the value of the effective Planck mass. The effective Planck mass $\chi$ has increased from a value when density fluctuations have left the horizon,
\be\label{38}
\bar\chi^2=m^2\bar x=\frac{\tau m^2}{40},
\ee
to the present value $\chi^2_0=M^2$. 

The dimensionful parameter in the cosmon kinetic term $K$ sets a characteristic scale of our model,
\be\label{39}
\bar m=\sqrt{12\tau m^2}.
\ee
A second characteristic scale corresponds to $\mu$, as given by eq. \eqref{9}. The parameter $\mu^{-1}$ is therefore set by the present horizon. One could choose units with $\mu=1$ and only discuss ratios of mass scales. The other dimensionful parameter is somewhat larger than $\mu$, cf. eq. \eqref{32A}
\be\label{39a}
\bar m=\sqrt{\frac{12\tau m^2}{\mu^2}}\mu=\sqrt{12}\cdot 10^5\mu=7\cdot 10^{-28}eV.
\ee
Both fundamental scales are much smaller than all known energy scales in particle physics. This is due to the very large value of the present Planck mass $M=\chi(t_0)$ in units of $\mu$, which is, in turn, the result of a long cosmological evolution,
\be\label{40}
\frac M\mu\approx 10^{60}.
\ee
All present masses of elementary particles are proportional to $M$, albeit often with a small proportionality constant of $10^{-16}$ or smaller. 

\medskip\noindent
 {\em Cosmon-Higgs doublet interactions.}
\smallskip

We finally discuss the coupling of the cosmon to the particles of the standard model. They are responsible for the production of the entropy and the heating of the universe at the end of the inflationary period.
We concentrate on the couplings of the cosmon to the Higgs doublet $\tilde h$. They may be responsible for transfering energy from the cosmon to the Higgs sector, and thereby to all the standard model particles which couple to the Higgs scalar. (The important role of the Higgs doublet is a feature shared with the discussion in ref. \cite{GB}, but we will find also important differences.)

Not much is known about the cosmon-Higgs interactions during the inflationary period. This may be seen by discussing in the Jordan frame an effective potential of the form
\be\label{46A}
\tilde V_h=\frac12\lambda_h(\tilde h^\dagger\tilde h+\mu^2_h-\bar\epsilon_h\chi^2)^2.\ee
For the large values of $\chi^2$ relevant for late cosmology the term $\mu^2_h$ can be neglected as compared to $\bar\epsilon_h\chi^2$. From present observations we have therefore no knowledge about $\mu^2_h$. During inflation, however, one may have $|\mu^2_h|\gg|\bar\epsilon_h\chi^2|$. Then $\mu^2_h>0$ will induce a minimum of $V_h$ at $\tilde h=0$, while for $\mu^2_h<0$ the minimum is at $\tilde h^\dagger\tilde h=-\mu^2_h>0$. If $\mu^2_h-\bar\epsilon_h\chi^2$ changes from positive to negative values during inflation such a model realizes hybrid inflation. Furthermore, $\lambda_h$ may depend substantially on $\chi^2/\mu^2$, with larger values during inflation as compared to late cosmology.

Let us parameterize the effective
 Higgs potential (after inclusion of all quantum fluctuations) by
\be\label{41}
\tilde V_h=\frac12\lambda_h(\tilde h^\dagger\tilde h-\epsilon_h\chi^2)^2.
\ee
Here, the dimensionless coupling functions  $\lambda_h$ and $\epsilon_h$ depend on the ratios $\tilde h^\dagger\tilde h/\chi^2$ and $\chi^2/\mu^2$. We will argue that the effective values that these functions take for
values of $\chi$ corresponding to 
 the inflationary phase may strongly differ from the present values. Large values of $\epsilon_h$ and $\lambda_h$ at the end of inflation facilitate an efficient entropy production and heating of the Universe. Unfortunately, the predictivity which could result from using for inflation
 the parameters for large $\chi$, namely those measured by experiments in the present cosmological epoch, is lost.  

Asymptotic scale invariance for $\chi^2/m^2\to\infty$ requires that $\lambda_h$ and $\epsilon_h$
 do no longer depend on explicit mass scales in this limit, and we define
\ba\label{42}
\bar\lambda_h(\tilde h^\dagger\tilde h/\chi^2)&=&\lambda_h(\tilde h^\dagger\tilde h/\chi^2,\chi^2/m^2\to\infty),\nn\\
\bar\epsilon_h(\tilde h^\dagger\tilde h/\chi^2)&=&\epsilon_h(\tilde h^\dagger\tilde h/\chi^2,\chi^2/m^2\to\infty).
\ea
For fixed $\chi^2$ and varying $\tilde h^\dagger\tilde h$ these couplings are still running in the standard way, due to the coupling of the Higgs boson to the fermions and gauge bosons of the standard model. For late cosmology we can replace $\lambda_h$ and $\epsilon_h$ by the scaling limit \eqref{42}. 

The dependence of $\bar\lambda_h$ and $\bar\epsilon_h$ on $\tilde h^\dagger \tilde h/\chi^2$ is governed by the $\beta$-functions of the effective theory for particles with masses below $\chi$. We may assume that this effective theory is the standard model of particle physics. In particular, the 
 running of $\bar\epsilon_h$ is governed by an anomalous dimension
\be\label{43}
\partial_t\bar\epsilon_h=A_h\bar\epsilon_h~,~t=\frac12\ln (\tilde h^\dagger\tilde h/\chi^2).
\ee
This particular form of the flow equation is due to an effective ``low energy dilatation symmetry'' associated to the second order character of the electroweak phase transition at $\bar\epsilon_h=0$. The low energy dilatation symmetry protects small values of $\bar\epsilon_h$, rendering them ``technically natural'' \cite{CWgaugehierarchy,F4,Bar,Ni1,Ni2}. The electroweak gauge hierarchy requires a tiny value $\bar\epsilon_h\approx 10^{-32}$. This has to hold for $\tilde h^\dagger\tilde h/\chi^2\approx 10^{-32}$. Due to the small value of $A_h$ (proportional to squared gauge and Yukawa couplings) a tiny value is also required for larger values of $\tilde h^\dagger\tilde h/\chi^2$. The running of $\bar\lambda_h$ obeys the usual flow equations in the standard model. A fixed point for quantum gravity or some other mechanism may impose $\bar\lambda_h(\tilde h^\dagger\tilde h/\chi^2=1)\ll 1$, resulting in the prediction for the mass of the Higgs boson $M_H\approx 126$ GeV \cite{SW,wheretolook}.

In the scaling limit the (partial) minimum of the Higgs potential occurs for $\tilde h_0\sim \chi$, as determined by
\be\label{44}
\tilde h_0^\dagger\tilde h_0=\bar\epsilon_{h,0}\chi^2,\quad
\bar\epsilon_{h,0}=\bar \epsilon_h(\tilde h^\dagger_0\tilde h_0/\chi^2).
\ee
Therefore the Fermi scale and associated particle masses depend on time, being always proportional to $\chi$. In consequence, dimensionless ratios, as electron mass over Planck mass, are time independent despite the time dependence of both the Planck mass and the Fermi scale. We also assume a constant value of the gauge couplings at some sliding scale $M_U\sim \chi$, for example the scale of grand unification, $\bar\alpha_s(M_U)=$const., and similar for Yukawa couplings. Then the 
gauge and Yukawa couplings, evaluated at a fixed ratio $\tilde h_0^\dagger \tilde h_0/\chi^2$, remain constant despite the variation of $\chi$. In consequence, all the 
severe observational bounds on the time dependence of fundamental couplings are met. (For possible small deviations from exact scaling and their observational consequences through a time variation of couplings and apparent violations of the equivalence principle see ref. \cite{CW1,CWcoupl}.) 

For the inflationary period the scaling limit $\chi^2/m^2\to\infty$ does not yet apply. (This holds despite the fact that for large $\tau$ the ratio $x=\chi^2/m^2$ is already substantially larger than one.) Let us assume that the $x$-dependence of $\lambda_h$ and $\epsilon_h$ is generated by flow equations of the type
\ba\label{45}
y\frac{\partial}{\partial y}\lambda_h&=&-B_\lambda(\lambda_h-\bar\lambda_h),\nn\\
y\frac{\partial}{\partial y}\epsilon_h&=&-B_\epsilon(\epsilon_h-\bar\epsilon_h)~,~y=\frac{\chi^2+\mu^2}{\mu^2}.
\ea
The fixed points at $\bar{\lambda}_h$ and $\bar{\epsilon}_h$ guarantee the asymptotic behaviour \eqref{42}.
(The positive ``anomalous dimensions'' $B_\lambda$ and $B_\epsilon$ can depend on $\tilde h^\dagger\tilde h/\chi^2$.) For the approximation of constant $B_\lambda$ and $B_h$ the solution of eq. \eqref{45} reads
\ba\label{46}
\lambda_h&=&\bar\lambda_h+\lambda^0_h\left(\frac{\chi^2+\mu^2}{\mu^2}\right)^{-B_\lambda},\nn\\
\epsilon_h&=&\bar\epsilon_h+\epsilon^0_h\left(\frac{\chi^2+\mu^2}{\mu^2}\right)^{-B_\epsilon}.
\ea
In particular, we observe that for $B_\epsilon\gtrsim 1/4$ one could have $\epsilon_h(x=0)=\epsilon_h^0\approx 1$. The flow from $\chi^2\approx \mu^2$ to $\chi^2=M^2$ would realize in this case an early phase of attraction to an (approximate) fixed point at $\epsilon_h\approx 0$ in the spirit of ref. \cite{strongint,wheretolook}. (Note that $\epsilon_h$ should reach $\bar\epsilon_h$ before nucleosynthesis.) 

At present, we have no knowledge on $B_\lambda,B_\epsilon$ or $\lambda^0_h,\epsilon^0_h$. The flow \eqref{45} is dominated by fluctuations with momenta of the order $\chi$. It may arise from some unified theory which becomes valid at the Planck scale, possibly in higher dimensions. Alternatively, it could result from the behavior near an ``ultraviolet fixed point'' defining quantum gravity. Also the simple effective potential \eqref{46A} may be cast (approximately) in this  form, with $B_\epsilon=-1$ and $\mu^2_h=-\epsilon^0_h\mu^2$. We conclude that the relevant interactions between the cosmon and the Higgs doublet during inflation are unknown and can be chosen freely as long as no information from an ``ultraviolet completion'' becomes available. For example, one
may assume that for $\chi^2=\mu^2$ no particularly small or large parameters appear in the potential, with $\lambda^0_h$ and $\epsilon_h^0$ of the order one. 

\medskip\noindent
 {\em Entropy production and heating.}
\smallskip

The entropy production at the end of inflation is best discussed in the familiar Einstein frame.
The Weyl scaling \eqref{4} results in a multiplication of $\tilde V_h$ by a factor $w^4$, such that $V_h$ obtains from $\tilde V_h$ by replacing in eq. \eqref{41} $\lambda_h$ by $\lambda'_h$,
\be\label{46AA}
\lambda'_h=\frac{\lambda_h M^4}{(\chi^2+m^2)^2}.
\ee
On the other hand, a standard kinetic term $\partial^\mu\tilde h^\dagger\partial_\mu\tilde h$ in the Jordan frame becomes in the Einstein frame
\ba\label{46B}
{\cal L}^{\rm kin}_h&=&\frac{M^2}{\chi^2+m^2}\partial^\mu\tilde h^\dagger\partial_\mu 
\tilde h=\partial^\mu h^\dagger\partial_\mu h\\
&&+\frac{\chi^2h^\dagger h}{(\chi^2+m^2)^2}\partial^\mu\chi\partial_\mu\chi+
\frac{\chi\partial^\mu\chi}{\chi^2+m^2}\partial_\mu(h^\dagger h),\nn
\ea
with
\be\label{46C}
h=\frac{M}{\sqrt{\chi^2+m^2}}\tilde h.
\ee
Expressed in terms of $h$ the effective potential for the Higgs doublet reads
\ba\label{46D}
V_h&=&\frac12\lambda_h\left(h^\dagger h-\epsilon_h\frac{\tilde\sigma^2}{4+\tilde \sigma^2/M^2}\right)^2\nn\\
&=&\frac12 \lambda_h\left(h^\dagger h-\frac{\epsilon_h M^2}{1+m^2/\chi^2}\right)^2.
\ea
In late cosmology one can neglect $m^2/\chi^2$ and the minimum of $V_h$ occurs in the Einstein frame at a fixed value $h^\dagger h=\epsilon_h M^2$. Corrections to the standard kinetic term for $h$ are suppressed by $h^\dagger h/\chi^2$ and negligible.

For the inflationary epoch the couplings between $h$ and $\chi$ are more complicated.
At the end of the inflationary phase we can neglect $\bar\epsilon_h$ and take $\chi^2\gg\mu^2$. 
For $\epsilon^0_h>0$ the 
 partial minimum of the potential \eqref{46D} with respect to $h$ obeys
\be\label{47}
h^\dagger h=\epsilon^0_h\left(\frac{m^2\tilde\sigma^2}{4\mu^2M^2}+1\right)^{-B_\epsilon}
\frac{\tilde\sigma^2}{4+\tilde \sigma^2/M^2},
\ee
defining a ``valley'' in the landscape of the potential. During the slow roll period of inflation the cosmological value of the Higgs doublet $h(t)$ follows this valley. A description in terms of coherent scalar fields remains valid. For sufficiently large ${\epsilon_h^0}$ and ${\lambda_h^0}$ the mass of the Higgs doublet around the time of horizon crossing of the density fluctuations is sufficiently large such that the single field inflation described previously remains valid. Coherent oscillations of $h$ around the valley could result in a periodic signal in the fluctuations.

The slow motion along the valley remains no longer valid after the end of inflation. A fast change of $\tilde{\sigma}$ can induce substantial oscillations of $h$. Non-equilibrium processes result finally in the production of incoherent fluctuations of the Higgs fields, or equivalently, the production of an incoherent density of Higgs particles. In turn, the Higgs particles produce fermions and gauge bosons, producing in this way the entropy of the universe. The plasma of standard model particles is interacting sufficiently strongly such that thermal equilibrium can be established. At the end of this ``heating of the universe'' the cosmology is described by a radiation dominated Friedman universe,  with an additional small part of early dark energy according to the scaling solution with $\Omega_h=nk^2_\infty$. 

For the alternative $\epsilon^0_h<0$ the potential minimum lies at $h=0$ as long as $\chi^2/\mu^2<|\epsilon^0_h/\bar\epsilon_h|^{1/B_\epsilon}$ (or $\chi^2/\mu^2_h<\bar\epsilon_h$ for eq. \eqref{46A}), changing to a non-zero value for larger $\chi$. Again a single field inflation is a good approximation if the doublet mass is large enough at the time of horizon crossing, with a possible periodic imprint from oscillations of $h$ around zero. If the potential minimum moves away from zero towards the end of inflation the model realizes hybrid inflation, with the possibility of an efficient entropy production.

Details of entropy production and heating depend on the parameters of the model. One useful point of view is the decay of cosmons  into Higgs particles. The latter is directly induced by a cubic coupling $\gamma_h\chi h^\dagger h$ which obtains by taking a derivative of the effective action with respect to $\chi,h$ and $h^\dagger$. This cubic coupling depends on the cosmological background field $\chi$. It can be large (in units of $\chi$) for large enough $\lambda_h^0\epsilon_h^0$.

The coupling of the cosmon to the Higgs boson is only one of the possibilities for entropy production. The cosmon may couple as well to other scalar fields, as the ones responsible for spontaneous symmetry breaking of a grand unified symmetry. Furthermore, particle production could proceed during an epoch when the kinetic energy of the cosmon dominates \cite{Fo,Spo,PV}. In view of the rich and diverse possibilities and the unknown parameters at the end of inflation it is hard to make any predictions for the heating of the Universe. A very rapid entropy production and effective thermalization or a delayed one seem both possible. 

We finally should mention that we  have assumed implicitly that the influence of the Higgs doublet on the cosmon is small at the time when the primordial density fluctuations cross the horizon. This justifies that we also neglect a possible interaction $\sim \xi_h\tilde h^\dagger\tilde h\tilde R$ \cite{CW3,Sh}. (In a different setting, and for very large $\xi_h$, this term plays a crucial role in Higgs inflation \cite{Sh} and Higgs-dilaton inflation \cite{SZ1,SZ2,BS}.)

\medskip\noindent
 {\em Unified picture and predictivity.}
\smallskip

In conclusion, we have found a rather simple setting for which the inflaton can be identified with the cosmon. While the cosmon-potential can be dominated by the same term $\tilde V=\mu^2\chi^2$, a change in the scalar kinetic term from large $K$ (inflation) to small $K$ (scaling solution) can explain the end of inflation with a transition to the radiation dominated epoch. This class of models is predictive, leading to a spectral index $n=0.97$ and an amplitude ratio for tensor to scalar fluctuations $r=0.13$. However, parameter values with smaller tensor amplitude $r$ are possible as well.  

One may ask if some of the parameters which are relevant for the behavior of early dark energy during the scaling solution can be connected to properties of inflation. This seems not to be the case in this class of models. The relevant parameter $k_\infty=1/\alpha$ which fixes the scaling ratio of early dark energy $\Omega_{h,e}$ plays only a subdominant role during the inflationary stage. The parameters characterizing the present properties of dark energy, as the dark energy fraction $\Omega_{h,0}\approx 0.7$ or the equation of state $w_{h,0}$ will be mainly determined by the mechanism which stop the evolution of the cosmon at the end of the matter dominated period. Here the connection to the parameters which are relevant for the inflationary period seems even weaker. 

For $\tilde V=\mu^2\chi^2$ present dark energy density fixes $\mu=2\cdot 10^{-33}$eV and therefore the value of the scalar potential for every value of $\chi$. The fluctuation amplitude $\Delta$ depends, however, on the value $\chi_l$ when a fluctuation of a given length scale $l$ has left the horizon. In turn, $\chi_l$ depends strongly on the parameter $\bar m$ in the kinetic term, giving an additional dependence of $\Delta$ on $\bar m$. On the other side, the terms involving $\bar m$ are completely irrelevant for late cosmology and can therefore not be extracted from the present background cosmology. The spectral index and the scalar to tensor ratio are found to be independent of $\bar m$ or $k_\infty$. This is the reason for the increased predictivity for these quantities for this class of models. On the other hand, the connection to late cosmology is not necessary for these predictions. 
For more general models the connection between $n,r$ and $\Delta$ on one side and the parameters determining the present cosmological evolution seems even weaker.

The unified picture of inflation and late Dark Energy is conceptually very appealing. Only one single cosmon field is needed for both epochs. At the moment no testable predictions of this connection are visible. This will have to wait until a more fundamental setting, presumably related to quantum gravity, further restricts the free parameters of the model.

\bibliography{cosmon_inflation}

\end{document}